# Analysing high-throughput sequencing data in Python with HTSeq 2.0

Givanna H Putri [1,2], Simon Anders [3], Paul Theodor Pyl [4], John E Pimanda [1,2,5,6] and Fabio Zanini [1,2,7,+]

[1] Prince of Wales Clinical School, University of New South Wales, Sydney, NSW, Australia
[2] Adult Cancer Program, Lowy Cancer Research Centre, University of New South Wales, Sydney, NSW, Australia
[3] Bioquant Center, University of Heidelberg, 69120 Heidelberg, Germany
[4] Division of Surgery, Oncology and Pathology, Department of Clinical Sciences Lund, Faculty of Medicine, Lund University, Lund, Sweden
[5] Department of Pathology, School of Medical Sciences, University of New South Wales, Sydney, NSW, 2052, Australia
[6] Department of Haematology, the Prince of Wales Hospital, Sydney, NSW, Australia
[7] Cellular Genomics Futures Institute, University of New South Wales, Sydney, NSW, Australia

[+] To whom correspondence should be addressed.

## Abstract

**Summary**: *HTSeq 2.0* provides a more extensive API including a new representation for sparse genomic data, enhancements in *htseq-count* to suit single cell omics, a new script for data using cell and molecular barcodes, improved documentation, testing and deployment, bug fixes, and Python 3 support.
**Availability and implementation:** HTSeq 2.0 is released as an open-source software under the GNU General Public Licence and available from the Python Package Index at https://pypi.python.org/pypi/HTSeq. The source code is available on Github at https://github.com/htseq/htseq.
**Contact**: fabio.zanini@unsw.edu.au

## Text

Single cell omics have exploded in popularity over the last few years, spearheaded by single cell transcriptomics. While commercial software solutions from manufacturers such as 10X Genomics and BD Biosciences provide standardized pipelines (e.g.

*cellranger*) for analysing single cell omics data, numerous experimental approaches rely on open source software to align reads and subsequently to quantify biological phenomena such as gene expression, chromatin accessibility, transcription factor binding affinities, and three-dimensional chromatin conformation. *HTSeq* (Anders, Pyl, and Huber 2015) was initially developed as a general purpose tool to analyse high-throughput sequencing data in Python. In parallel, the *htseq-count* script was designed to count the number of reads or read pairs attributable to distinct genes in bulk RNA-Seq experiments. At that time, single cell approaches were limited to specialized biotechnology laboratories. In this application note, we report the development of *HTSeq 2.0*, which improves the general-purpose application programming interface (API) and specifically *htseq-count* to encompass diverse omics analyses, including single cell RNA sequencing (scRNA-Seq).

First, we have improved *htseq-count*, a popular script used to quantify gene expression in bulk and scRNA-Seq experiments (**Figure 1A-C**). Multiple BAM files can now be processed with a single call of the script, which results in a counts table with each row or column representing the counts from a separate BAM file. This is not only convenient but also faster because genomic features are loaded only once from the GTF file, which can take as long as processing the reads for a typical plate-based single cell experiment. If multiple cores are available on the machine, *htseq-count* is now able to parallelize the quantification by allocating distinct input BAM files to each core (**Figure 1A**). The script also supports more output formats: compressed sparse matrices via *scipy* (Virtanen et al. 2020), *mtx* files in the style of *cellranger*, and h5-like file formats such as *h5ad* (Wolf, Angerer, and Theis 2018) and *loom* (http://loompy.org) (**Figure 1A**). These output formats make it easier for users to import the counts table into downstream analysis libraries, especially single cell ones such as *scanpy* (Wolf, Angerer, and Theis 2018) and *singlet* (https://github.com/iosonofabio/singlet). We also added support for storing additional metadata for each genomic feature. This has two clear applications: (i) Tracking additional gene information such as chromosome or aliases, which is useful for instance to exclude sex chromosomes in downstream analyses, and (ii) Collecting disaggregated exon-level counts, which provides a simple yet powerful approach to quantifying differential isoform expression (**Figure 1B**). To encourage users to customize their analysis pipeline, we also restructured the key steps of *htseq-count* into well documented functions and added a tutorial that explains feature counting step by step. Additionally, through a new script called *htseq-count-barcodes,* we support quantification of features in data multiplexed via cell barcodes and unique molecular identifiers (UMIs). Among other applications, the new script enables custom re-analysis of BAM files produced by *cellranger* with different parameters.

One of the key data structures in *HTSeq* has been *StepVector*, an efficient sparse representation for piecewise-constant values on a one-dimensional discrete space (typically a chromosome) (**Figure 1D**). As an example, it can be used to store overlaps between gene bodies, critical for removing ambiguities in downstream gene expression analyses. However, genomic data is sometimes characterized by a distinct type of sparsity whereby the data appears as dense "islands of knowledge" in a sea of missing data. This type of sparsity is apparent in the read coverage produced by amplicon sequencing or ChIP-Seq where most of the genome is uncovered, but nonzero rapidly fluctuating coverage, down to a single nucleotide resolution (e.g. due to single nucleotide variations), are present only around specific kilobase-long stretches. To represent this type of sparsity efficiently, we created a new class called *StretchVector*. At its core, a *StretchVector* is a collection of stretches implemented via dense numpy arrays (Harris et al. 2020), each with associated start-end coordinates (**Figure 1E**). Each stretch represents an island of data, while the rest of the genome is not stored. We implemented functions for stretch extension, trimming, resetting, shifting, views or slices, copy, and lastly conversion to and from monolithic arrays for simple data ingestion/extraction. Separately from *StretchVector*, we also improved the support for custom ChIP-Seq and chromatin conformation capture (high C) analyses by adding parsers for bedGraph and BigWig files via *pyBigWig* (Ryan et al. 2021) and by writing new dedicated tutorials.

Lastly, we improved the API of *HTSeq* as a whole and made architectural changes to the package to ensure its compatibility with current software development standards. Among other things, we (1) modernized the codebase to Python 3, (2) added provisions for continuous integration and development including automatic binary releases on multiple architectures, (3) established unit tests and test suites, (4) fixed bugs, and (5) added support for improved dependency infrastructure such as autodetection of SAM/BAM/CRAM file type via *HTSlib* (Bonfield et al. 2021). All aforementioned changes were carried out without compromising the efficiency of *HTSeq*, which stems from a cross-language design via Cython (Behnel et al. 2011) and SWIG (Beazley 2003).

In conclusion, *HTSeq* 2.0 is a fast and reliable Python library for not only analysing high-throughput sequencing data, but also for quantifying gene expression from bulk and single cell RNA-Seq experiments. Compared to the previous implementation, we added specific support for single cell experiments and a richer API including a new class for "islands-of-data" sparsity, improved API documentation and tutorials, fixed a number of bugs, and established a robust testing and deployment framework to ensure scientific reproducibility and enable continuous code integration. We believe these improvements will make *HTSeq* 2.0 a convenient tool for exploring and quantifying high-throughput sequencing experiment results across multiple omic modalities.

## Acknowledgements

We would like to thank Wofgang Huber for scientific exchanges and all HTSeq contributors for their valuable time. This work was partially funded through a long-term European Molecular Biology Organization Fellowship ALTF 269–2016 to F.Z. and through GRANT GNT1200271 from the National Health and Medical Research Council to J.E.P.

## Figure

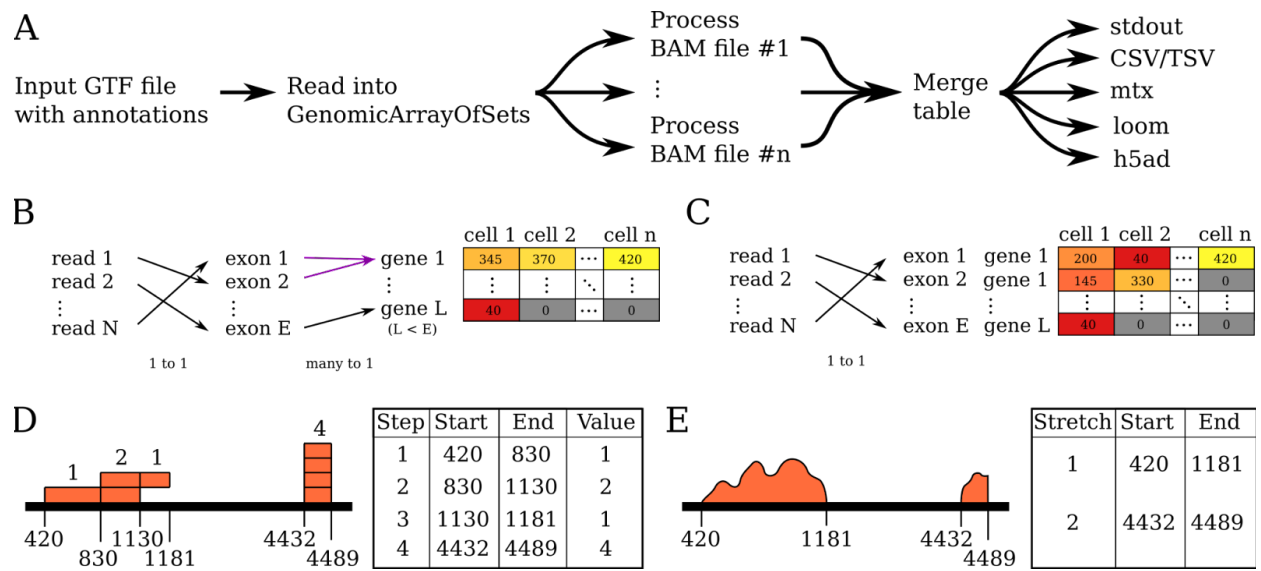

**Figure 1: Major HTSeq 2.0 improvements.** (**A-C**): Improvements to *htseq-count*. (**A**) Parallel processing on multicore architectures enables faster processing of single cell data, where each cell is represented by a BAM file (as is typically the case for Smart-seq2 (Picelli et al. 2013) and viscRNA-Seq (Zanini et al. 2018)). Note the new output formats available in HTSeq 2.0. (**B**) Conventional gene-cell matrix, which collapses reads that align to distinct exons of the same gene into a single gene count. (**C**) Additional attributes enable quantification at the exon level while retaining information on which gene each exon belongs to. (**D-E**): Sparse data representations in *HTSeq 2.0*. (**D**) *StepVector* represents piecewise constant sparse genomic data. (**E**) *StretchVector* represents sparse islands of genomic data instead.